\RequirePackage{ifpdf}
\ifpdf 
\documentclass[pdftex]{sigma}
\else
\documentclass{sigma}
\fi

\begin{document}

\renewcommand{\PaperNumber}{***}

\FirstPageHeading

\ShortArticleName{Characteristic Lie rings and symmetries of differential Painlev\'e I and Painlev\'e III equations}

\ArticleName{Characteristic Lie rings and symmetries of differential Painlev\'e I and Painlev\'e III equations}

\Author{Olga S. KOSTRIGINA~$^\dag$ and Anatoliy V. ZHIBER~$^\ddag$}

\AuthorNameForHeading{O.S.~Kostrigina and A.V.~Zhiber}

\Address{$^\dag$~Ufa State Aviation Technical University, 12, K. Marx str., Ufa, Russia,} 
\EmailD{\href{mailto:email@address}{o\_kostrigina@mail.ru}} 

\Address{$^\ddag$~Institute of Mathematics, RAS, 112, Chernyshevsky str., Ufa, Russia}
\EmailD{\href{mailto:email@address}{zhiber@mail.ru}} 


\ArticleDates{Received ???, in final form ????; Published online ????}

\Abstract{Two-component hyperbolic system of equations generated by ordinary differential Painlev\'e I 
\[
u_{yy}=6u^2+y
\]
and Painlev\'e III
\[
yuu_{yy}=yu^2_{y}-uu_y+\delta y+\beta u+\alpha u^3 +\gamma yu^4
\] 
equations are considered, where $\alpha, \beta, \gamma, \delta$ are complex numbers. 
The structure of characteristic Lie rings is studied and higher symmetries of Lie-B\"acklund are obtained.}


\Classification{35L70} 

\section{Introduction}

Characteristic Lie rings and higher symmetries of Lie-B\"acklund for Painlev\'e I 
\begin{equation}
u_{yy}=6u^2+y\label{P1}
\end{equation}
and Painlev\'e III 
\begin{equation} 
yuu_{yy}=yu^2_{y}-uu_y+\delta y+\beta u+\alpha u^3 +\gamma yu^4\label{penleve3}
\end{equation}
equations are considered, where $\alpha, \beta, \gamma, \delta$ are complex numbers.

For definition of characteristic Lie ring of ordinary differential equations consider hyperbolic system of eqautions (see \cite{ZHIBER4})
\begin{equation}
u^i_{xy}=F^i(x,y,u, u_x, u_y),\  \ i=1,2,\ldots, n, \ u=(u^1,u^2,\ldots,u^n).\label{main_eq1}
\end{equation}
We introduce a set of independent variables $u_1=u_x,\ \bar{u}_1=u_y,\ u_2=u_{xx},\ \bar{u}_2=u_{yy},\ldots$ and denote by $ \bar{D} (D)$ an operator of full differentiation with respect to $y\ (x)$.
\begin{definition}
Function $\omega=\omega(x,y,u,u_1,\ldots,u_m)$ is called an $x-$integral of order $m$ of equations \eqref{main_eq1} if $\displaystyle\sum\limits_{i=1}^{n} \left( \frac{\partial \omega}{\partial u^{i}_{m}} \right)^{2} \neq 0\  \mbox{and} \ \bar{D}\omega=0.$ Similarly, $\bar{\omega}=\bar{\omega}(x,y,u,\bar{u}_1,\ldots,\bar{u}_p)$ is an $y-$integral of order $p$ of equations \eqref{main_eq1}, if $\displaystyle\sum\limits_{i=1}^{n} \left( \frac{\partial \bar{\omega}}{\partial \bar{u}^{i}_{p}} \right)^{2} \neq 0\  \mbox{and} \ D\bar{\omega}=0.$
\end{definition}
Denote an area of locally analytical functions by $\Im$, where every function depends on a finite number of variables $x,y,\bar{u}_1, u, u_1, u_2,\dots,u_k,\ldots$. Operator $\bar{D}$ acts on functions from $\Im$ as follows:
\[
\bar{D}=\bar{u}_2^iX_i+X_{n+1},
\]
where
\begin{equation*}
\begin{split}
&\qquad \qquad \qquad \qquad  X_i=\frac{\partial}{\partial \bar{u}_1^i}, \ i=1,2,\ldots,n,  \\
&X_{n+1}=\frac{\partial}{\partial y}+\bar{u}_1^i\frac{\partial}{\partial u^i}+F^i\frac{\partial}{\partial u_1^i}+D(F^i)\frac{\partial}{\partial u_2^i}+\ldots +D^{k-1}(F^i)\frac{\partial}{\partial u_k^i}+\ldots.
\end{split}
\end{equation*}

The $x-$characteristic Lie ring of equations (\ref{main_eq1}) is a ring $A$ generated by vector fields\\ $X_1, X_2,\ldots ,X_{n+1}$. Similarly, the $y-$characteristic Lie ring $\bar{A}$ is defined.

Notice that the concept of characteristic vector field first was introduced by the Goursat in \cite{goursat}.

Denote a linear area of commutator of length $n, \ n=1,2,3,\ldots$ by $L_n.$ For example $L_1$ is a linear span of vector fields $X_1,X_2,\ldots,X_{n+1}$ and $L_2$ is generated by operators $X_{ij}=[X_i,X_j], \ i,j=1,2,\ldots,n+1$ etc. Then the $x-$ring is represented in the form
\[
A=\sum^{\infty}_{i=1} L_i.
\] 
For $y-$characteristic ring $\bar{A}$ we have 
\[
\bar{A}=\sum^{\infty}_{i=1} \bar{L}_i.
\]

In \cite{murtazina} it was conjectured that dimension of linear areas $L_i \ (\bar{L}_i)$ for integrable equations grows slowly. Further this hypothesis was confirmed by numerous examples of integrable continuous and discrete models. Property of minimal growth became a kind of classification criterion for integrable equations. From \cite{umj} it is follows that this property of the ring and the property of existence of hierarchy of higher symmetries for integrable equations are equally universal.

\begin{definition}
Symmetry of equations (\ref{main_eq1}) is a set of functions 
\[
f^i=f^i(x,y,u,u_1,\ldots,u_r,\bar{u}_1,\ldots,\bar{u}_s), \ i=1,2,\ldots,n,
\]
satisfying the defining equations 
\[
D\bar{D}f^i=(F^i_{u^j}+F^i_{u_1^j}D+F^i_{\bar{u}_1^j})f^j, \ i=1,2,\ldots,n.
\]
\end{definition}

It is known that any Lie-B\"acklund symmetry $f^i, \ i=1,2,\ldots,n$ is represented in the form
\[
f^i=f^i_1+f^i_2,
\]
where functions $f^i_1$ are depends on variables $x,y,u,u_1,u_2,\ldots,$ and functions $f^i_2$  are depends on variables $x,y,u,\bar{u}_1,\bar{u}_2,\ldots.$ And besides $f^i_1$ and $f^i_2$ are symmetries itself.

Concept of characteristic Lie ring for a system of ordinary differential equations 
\begin{equation}\label{main_eq2}
u^i_{y}=f^i(x,y,u),\  \ i=1,2,\ldots, n, \ u=(u^1,u^2,\ldots,u^n),
\end{equation}
was introduced in \cite{gurg}, \cite{book}. In this work two definitions for characteristic ring were proposed (\ref{main_eq2}). First definition is based on change 
\[
u^i=\frac{\partial p^i}{\partial x}, \ i=1,2,\ldots, n,
\]
in which equations (\ref{main_eq2}) are take the form
\begin{equation}
p^i_{xy}=f^i(x,y,v_x),\  \ i=1,2,\ldots, n, \ v=(v^1,v^2,\ldots,v^n).\label{main_eq3}
\end{equation}
The $x$ and $y-$characteristic Lie rings of hyperbolic system (\ref{main_eq3}) are called characteristic Lie rings of initial system of ordinary differential equations (\ref{main_eq2}).

Another definition of characteristic Lie ring for system (\ref{main_eq2}) is associated with the following hyperbolic system of equations
\begin{equation}
u^i_{xy}=f^i_x+f^i_{u^k}u^k_x, \ i=1,2,\ldots, n.\label{main_eq22}
\end{equation}

In the first section characteristic Lie ring for system of equations (\ref{main_eq3}) and (\ref{main_eq22})\\ corresponding to Painlev\'e I equation are considered. We show that these Lie rings are rings of slow growth. Namely, for the system of equations (\ref{main_eq3}) dimension of $x-$ring is equal to three, and for $y-$ring $dim\bar{L}_1=3, \ dim\bar{L}_n=n,\ n=2,3,4; \ dim\bar{L}_5\leq5.$ For the hyperbolic system (\ref{main_eq22}), corresponding to the equation Painlev\'e I, it is shown that dimension of $y-$ring is four, and for $x-$ring the following formalus are true $dimL_1=3, \ dimL_k=1, \ k=2,3,4,5,6; \ \ dim\sum^6_{i=1}L_i=8.$

In the second and third sections the higher symmetries of Lie-B\"acklund for system of equations (\ref{main_eq3}) corresponding to equations (\ref{P1}) and (\ref{penleve3}) are obtained .

\section{Lie rings for Painlev\'e I equation}

We rewrite equation (\ref{P1}) as 
\[
u_y=v,\ \ v_y=6u^2+y.
\]
Then the corresponding hyperbolic system of equations (\ref{main_eq3}) and (\ref{main_eq22}) take the form
\begin{equation}
p_{xy}=q_x, \ \ q_{xy}=6p^2_x+y \ (u=p_x,\ \ v=q_x)\label{Ps1}
\end{equation}
and
\begin{equation}
u_{xy}=v_x,\ \ v_{xy}=12uu_x.\label{Ps012}
\end{equation}

In this section we study the characteristic Lie ring of systems of equations (\ref{Ps1}) and (\ref{Ps012}).

For the system (\ref{Ps1}) $x-$characteristic Lie ring is generated by vector fields 
\begin{gather*}
X_1=\frac{\partial}{\partial p}, \ \ X_2=\frac{\partial}{\partial q}, \\
X_3=\frac{\partial}{\partial y}+q_1\frac{\partial}{\partial p_1}+(6p^2_1+y)\frac{\partial}{\partial q_1}+q_2\frac{\partial}{\partial p_2}+12p_1p_2\frac{\partial}{\partial q_2}+\ldots.
\end{gather*}
Since $[X_1,X_2]=[X_1,X_3]=[X_2,X_3]=0,$ then dimension of $x-$ring is equal to three. Herewith $x-$integrals
\[
\omega=\omega(y,p_1,q_1) \ \mbox{and}\ w=w(y,p_1,q_1)
\]
are defined from partial derivative equations 
\begin{equation}
\left(\frac{\partial}{\partial y}+q_1\frac{\partial}{\partial p_1}+(6p^2_1+y)\frac{\partial}{\partial q_1}\right)F=0.\label{w}
\end{equation}
Note, that $\omega =const$ and $w=const$ are define integrals of initial equation (\ref{P1}).

The $y-$characteristic Lie ring of equations (\ref{Ps1}) is defined by vector fields
\begin{gather}
Y_1=\frac{\partial}{\partial p_1}, \ \ Y_2=\frac{\partial}{\partial q_1},\notag \\
Y_3=\frac{\partial}{\partial x}+p_1\frac{\partial}{\partial p}+q_1\frac{\partial}{\partial \lambda_1}+(6p^2_1+y)\frac{\partial}{\partial \lambda_2}+(12p_1q_1+1)\frac{\partial}{\partial \lambda_3}+\notag\\
+(12q^2_1+72p^3_1+12yp_1)\frac{\partial}{\partial \lambda_4}+(360q_1p^2_1+12p_1+36yq_1)\frac{\partial}{\partial \lambda_5}+\label{Y1}\\
+(2160p^4_1+720p_1q^2_1+48q_1+576yp^2_1+36y^2)\frac{\partial}{\partial \lambda_6}+\ldots\notag,
\end{gather}
where
\[
\frac{\partial}{\partial \lambda_i}=\frac{\partial}{\partial \bar{p}_i}+\frac{\partial}{\partial \bar{q}_{i-1}}, \ i=1,2,\ldots.
\]
It is easy to see, that the vector fields $Y_1,\ [Y_1,Y_3],\ [Y_1,[Y_1,Y_3]],\ [Y_1,[Y_1,[Y_1,Y_3]]],\ldots$ are linearly independent, and hence $y-$ring is infinite. 

Since $D$ and $\bar{D}$ commute, we have
\begin{gather}
[\bar{D},D]F(x,y,p,q,p_1,q_1,\bar{p}_1,\bar{q}_1,\bar{p}_2,\bar{q}_2,\ldots)=\notag\\
=\bar{D}(p_2Y_1+q_2Y_2+Y_3)F-(p_2Y_1+q_2Y_2+Y_3)\bar{D}F=\label{DD}\\
=(p_2[Y,Y_1]+q_2[Y,Y_2]+[Y,Y_3]+q_2Y_1+12p_1p_2Y_2)F=0.\notag
\end{gather}
Here $Y$ is the operator of full differentiation with respect to $y$ in the area of functions depending on a finite set of variables $x,y,p,q,p_1,q_1,\bar{p}_1,\bar{q}_1,\bar{p}_2,\bar{q}_2,\ldots$
\[
Y=\frac{\partial}{\partial y}+\bar{p}_1\frac{\partial}{\partial p}+\bar{q}_1\frac{\partial}{\partial q}+q_1\frac{\partial}{\partial p_1}+(6p^2_1+y)\frac{\partial}{\partial q_1}
+\bar{p}_2\frac{\partial}{\partial \bar{p}_1}+\bar{q}_2\frac{\partial}{\partial \bar{q}_1}+\bar{p}_3\frac{\partial}{\partial \bar{p}_2}+\bar{q}_3\frac{\partial}{\partial \bar{q}_2}+\ldots.
\]

The following statement is true.
\begin{lemma}\label{lem1}
Let the vector field $\bar{Z}$ is given by
\begin{gather*}
\bar{Z}=\sum^{\infty}_{i=1}\left(\alpha_i\frac{\partial}{\partial \bar{p}_i}+\beta_i\frac{\partial}{\partial \bar{q}_i}\right),\\
\alpha_i=\alpha_i(y,p_1,q_1),\ \beta_i=\beta_i(y,p_1,q_1).
\end{gather*}
Then the equality $[Y,\bar{Z}]=0$ is true if and only if $\bar{Z}=0.$
\end{lemma}
\begin{proof} We have
\[
[Y,\bar{Z}]=\sum^{\infty}_{i=1}\left(Y(\alpha_i)\frac{\partial}{\partial \bar{p}_i}+Y(\beta_i)\frac{\partial}{\partial \bar{q}_i}\right)-\sum^{\infty}_{i=1}\left(\alpha_{i}\frac{\partial}{\partial \bar{p}_{i-1}}+\beta_{i}\frac{\partial}{\partial \bar{q}_{i-1}}\right)=0,
\]
that is $\alpha_1=0,$ $\beta_1=0, \ \alpha_{i+1}=Y(\alpha_i), \ \beta_{i+1}=Y(\beta_i), \ i=1,2,\ldots.$ Hence $\alpha_i=0,$ $\beta_i=0, i=1,2,\ldots$ and $\bar{Z}=0.$ 

\end{proof}

From (\ref{DD}) it is follows that
\begin{equation}
[Y,Y_1]=-12p_1Y_2,\ [Y,Y_2]=-Y_1, \ [Y,Y_3]=0.\label{YY}
\end{equation}

The linear area $\bar{L}_2$ is generated by operators $Y_{13}$ and $Y_{23}:$ 
\begin{gather}
 Y_{13}=\frac{\partial}{\partial p}+12p_1\frac{\partial}{\partial \lambda_2}+12q_1\frac{\partial}{\partial \lambda_3}+(216p^2_1+12y)\frac{\partial}{\partial \lambda_4}+(720p_1q_1+12)\frac{\partial}{\partial \lambda_5}+\notag\\+
(8640p^3_1+720q^2_1+1152yp_1)\frac{\partial}{\partial \lambda_6}+\ldots,\label{Y2}\\
Y_{23}=\frac{\partial}{\partial \lambda_1}+12p_1\frac{\partial}{\partial \lambda_3}+24q_1\frac{\partial}{\partial \lambda_4}+(360p^2_1+36y)\frac{\partial}{\partial \lambda_5}+(1440p_1q_1+48)\frac{\partial}{\partial \lambda_6}+\ldots,\notag
\end{gather}
Using the Jacobi identity, we obtain
\[
[Y,Y_{13}]=-[Y_3,[Y,Y_1]]+[Y_1,[Y,Y_3]],\ \ [Y,Y_{23}]=-[Y_3,[Y,Y_2]]+[Y_2,[Y,Y_3]],
\]
or, using (\ref{YY}), we have
\[
[Y,Y_{13}]=-[Y_3,-12p_1Y_2],\ \ [Y,Y_{23}]=-[Y_3,-Y_1].
\]
Thus, we have the following relations
\begin{equation}
[Y,Y_{13}]=-12p_1Y_{23},\ \ [Y,Y_{23}]=-Y_{13}.\label{YY23}
\end{equation}

The linear area $\bar{L}_3$ is generated by commutators $Y_{113},\ Y_{123}, \ Y_{213}, \ Y_{223}.$ Herewith
\[
Y_{213}=Y_{123}
\]
and
\begin{gather}
Y_{113}=12\frac{\partial}{\partial \lambda_2}+432p_1\frac{\partial}{\partial \lambda_4}+{720q_1}\frac{\partial}{\partial \lambda_5}+(25920p^2_1+1152y)\frac{\partial}{\partial \lambda_6}+\ldots,\notag\\
Y_{123}=12\frac{\partial}{\partial \lambda_3}+720p_1\frac{\partial}{\partial \lambda_5}+1440q_1\frac{\partial}{\partial \lambda_6}+\ldots,\label{Y3}\\
Y_{223}=24\frac{\partial}{\partial \lambda_4}+1440p_1\frac{\partial}{\partial \lambda_6}+\ldots.\notag
\end{gather}
As above, we can prove the following formulas
\begin{equation}
[Y,Y_{213}]=-Y_{113}-12p_1Y_{223},\ \ [Y,Y_{113}]=-12Y_{23}-24p_1Y_{123},\ \ [Y,Y_{223}]=-2Y_{123}.\label{YY223}
\end{equation}

The linear area $\bar{L}_4$ is generated by operators $Y_{1113},\ Y_{2113}, \ Y_{1213},\ Y_{2213}, \ Y_{1223},$ $Y_{2223},$  $ [Y_{13},Y_{23}],$ herewith
\[
Y_{1213}=Y_{2113}, \ Y_{1223}=Y_{2213}, \ [Y_{13},Y_{23}]=0
\]
and 
\begin{gather}
Y_{1113}=432\frac{\partial}{\partial \lambda_4}+51840p_1\frac{\partial}{\partial \lambda_6}+\ldots,\notag\\
Y_{2113}=720\frac{\partial}{\partial \lambda_5}+0\cdot\frac{\partial}{\partial \lambda_6}+\ldots,\label{Y4}\\
Y_{2213}=1440\frac{\partial}{\partial \lambda_6}\ldots,\ \ Y_{2223}=0\cdot\frac{\partial}{\partial \lambda_6}+\ldots.\notag
\end{gather}
We can show that
\begin{equation}
[Y,Y_{1113}]=-36p_1Y_{2113}-36Y_{123}, \ \ [Y,Y_{2213}]=-2Y_{2113}-12p_1Y_{2223}.\label{YY1113}
\end{equation}

From (\ref{Y1}), (\ref{Y2}), (\ref{Y3}) it is follows that $dim\bar{L}_1=3,\ dim\bar{L}_2=2,\ dim\bar{L}_3=3.$ Let us show that the operators $Y_{223},Y_{1113},Y_{2113},Y_{2213},Y_{2223}$ are linearly independent. If they are dependent then according to (\ref{Y3}), (\ref{Y4}) we must have
\[
Y_{1113}-\frac{432}{24}Y_{223}=\frac{25920}{1440}p_1Y_{2213}.
\]
According to Lemma \ref{lem1}, this relation is equivalent to
\[
[Y,Y_{1113}]-18[Y,Y_{223}]=18q_1Y_{2213}+18p_1[Y,Y_{2213}].
\]
Using the formulas (\ref{YY223}), (\ref{YY1113}), we have 
\[
-36p_1Y_{2113}-36Y_{123}+36Y_{123}=18q_1Y_{2213}+18p_1(-2Y_{2113}-12p_1Y_{2223})
\] 
or
\[
q_1Y_{2213}-12p_1^2Y_{2223}=0.
\] 
That it is impossible. Hence operators $Y_{223},Y_{1113},Y_{2113},Y_{2213},Y_{2223}$  are linearly independent and $dim\bar{L}_4=4.$

The area $\bar{L}_5$ is generated by commutators $Y_{11113},$ $Y_{21113},$ $Y_{12113},$ $Y_{22113},$ $Y_{12213},$ $Y_{22213},$ $Y_{12223},$ $Y_{22223},$ and among them there are pairs of equal operators. Namely $Y_{21113}=Y_{12113},\ Y_{22113}=Y_{12213}, \ Y_{22213}=Y_{12223}.$ Thus $dim\bar{L}_5\leq5.$ 

Based on the above results, it seems possible to assume the validity of the following formulas
\[
dim\bar{L}_k\leq k, \ k\geq6.
\] 
Then we have 
\[
dim(\bar{L}_1+\bar{L}_2+\ldots+\bar{L}_n)\leq3+2+3+4+5+\sum^{n}_{k=6}k,
\] 
or
\[
dim(\bar{L}_1+\bar{L}_2+\ldots+\bar{L}_n)\leq\frac{n^2}{2}+\frac{n}{2}+2.
\] 
Thus, we can hypothesize that the $y-$ring of (\ref{Ps1})  is a ring of slow growth. 

Next, we consider the characteristic Lie ring of equations (\ref{Ps012}).

$Y-$characteristic Lie ring of (\ref{Ps012}) is generated by operators
\begin{gather*}
Y_1=\frac{\partial}{\partial u_1}, \ Y_2=\frac{\partial}{\partial v_1},\\
Y_3=u_1\left(\frac{\partial}{\partial u}+12u\frac{\partial}{\partial \bar{v}_1}+12u\frac{\partial}{\partial \bar{u}_2}+12\bar{u}_1\frac{\partial}{\partial \bar{v}_2}+\ldots\right)+v_1\left(\frac{\partial}{\partial v}+\frac{\partial}{\partial \bar{u}_1}+12u\frac{\partial}{\partial \bar{v}_2}+\ldots\right).
\end{gather*}
Clearly that $Y_3=u_1Y_{13}+v_1Y_{23}$ and operators $Y_1, \ Y_2, \ Y_{13}, \ Y_{23}$ form a basis of $y-$ characteristic ring. Thus, the dimension of $y-$ring is equal to 4. And equations (\ref{Ps012}) has two $y-$integrals of the first order:
\[
\bar{\omega}=\bar{u}_1-v, \ \bar{w}=\bar{v}_1-6u^2.
\]

Consider $x-$characteristic ring. The operator of full differentiation with respect to $y$ in the area of functions depending on a set of variables $u,v,u_1,v_1,u_2,v_2,\ldots$ is given by
\[
\bar{D}=\bar{u}_1X_1+\bar{v}_1X_2+X_3,
\]
where
\begin{gather*}
X_1=\frac{\partial}{\partial u}, \ X_2=\frac{\partial}{\partial v},\\
X_3=\left(v_1\frac{\partial}{\partial u_1}+v_2\frac{\partial}{\partial u_2}+v_3\frac{\partial}{\partial u_3}+\ldots\right)+\\
+12\left(uu_1\frac{\partial}{\partial v_1}+(uu_2+u^2_1)\frac{\partial}{\partial v_2}+(uu_3+3u_1u_2)\frac{\partial}{\partial v_3}+\ldots\right).
\end{gather*}

Since operators $D$ and $\bar{D}$ commute, we have
\begin{gather}
[\bar{D},D]F(u,v,u_1,v_1,u_2,v_2,\ldots)=\notag\\
=(D(\bar{u}_1X_1+\bar{v}_1X_2+X_3)-(\bar{u}_1X_1+\bar{v}_1X_2+X_3)D)F=\label{DD1}\\
=(\bar{u}_1[D,X_1]+\bar{v}_1[D,X_2]+[D,X_3]+v_1X_1+12uu_1X_2)F=0.\notag
\end{gather}
Here $X$ is the operator of full differentiation with respect to $x$ in the area of functions depending on a set of variables $u,v,u_1,v_1,u_2,v_2,\ldots$
\[
X=u_1\frac{\partial}{\partial u}+v_1\frac{\partial}{\partial v}+u_2\frac{\partial}{\partial u_1}+v_2\frac{\partial}{\partial v_1}+\ldots.
\]

For the operator $X$  we can prove the following statement similarly as in the lemma \ref{lem1}.
\begin{lemma}\label{lem2}
Let the vector field $Z$ is given by
\begin{gather*}
Z=\sum^{\infty}_{i=1}\left(\delta_i\frac{\partial}{\partial u_i}+\epsilon_i\frac{\partial}{\partial v_i}\right),\\
\delta_i=\delta_i(u,v,u_1,v_1,u_2,v_2,\ldots,u_{n_i},v_{n_i}),\ \epsilon_i=\epsilon_i(u,v,u_1,v_1,u_2,v_2,\ldots,u_{k_i},v_{k_i}).
\end{gather*}
Then $[X,Z]=0$ if and only if $Z=0.$
\end{lemma}

From (\ref{DD1}) it is follows 
\begin{equation}
[X,X_1]=0,\ [X,X_2]=0, \ [X,X_3]=-v_1X_1-12uu_1X_2.\label{XX}
\end{equation}

It is easy to see that the dimension of the linear area $L_1 \ (L_1=L\left\langle X_1,X_2,X_3\right\rangle)$ is equal to three.

Since the coefficients of the vector field $X_3$ are independent of $v,$ then commutator $X_{23}=0.$ And commutator $X_{13}$ has the form
\[
X_{13}=12\left(u_1\frac{\partial}{\partial v_1}+u_2\frac{\partial}{\partial v_2}+u_3\frac{\partial}{\partial v_3}+u_4\frac{\partial}{\partial v_4}+\ldots\right).
\]
Thus $dimL_2=1.$ Herewith
\[
[X,X_{13}]=-[X_3,[X,X_1]]+[X_1,[X,X_3]],
\]
or, using (\ref{XX}), we have
\[
[X,X_{13}]=[X_1,-v_1X_1-12uu_1X_2],
\] 
that is
\begin{equation}
[X,X_{13}]=-12u_1X_2.\label{XX_13}
\end{equation}

The form of the operators $X_1, \ X_2, \ X_3$ and $X_{13}$  implies that  $X_{113}$ and $X_{213}$ are zero, and the commutator $X_{313}$ has the form
\begin{gather*}
X_{313}=12\left(v_1\frac{\partial}{\partial v_1}+v_2\frac{\partial}{\partial v_2}+v_3\frac{\partial}{\partial v_3}+v_4\frac{\partial}{\partial v_4}+\ldots\right)-\\
-12\left(u_1\frac{\partial}{\partial u_1}+u_2\frac{\partial}{\partial u_2}+u_3\frac{\partial}{\partial u_3}+u_4\frac{\partial}{\partial u_4}+\ldots\right).
\end{gather*}

Taking into account (\ref{XX}), (\ref{XX_13}), we have
\begin{gather*}
[X,X_{313}]=-[X_{13},[X,X_3]]+[X_3,[X,X_{13}]]=\\
=-[X_{13},-v_1X_1-12uu_1X_2]+[X_3,-12u_1X_2],
\end{gather*}
or
\begin{equation}
[X,X_{313}]=12u_1X_1-12v_1X_2.\label{XX_313}
\end{equation}

Let show that the operators $X_3, \ X_{13}$ and $X_{313}$  are linearly independent. Indeed, if they are linearly dependent, then there exists functions $\alpha, \ \beta$ of the variables $u,v,u_1,v_1,u_2,v_2,\ldots$ such that $X_{313}=\alpha X_3+\beta X_{13}.$  Last equality, according to Lemma \ref{lem2},  is equivalent to
\[
[X,X_{313}]=D(\alpha)X_3+D(\beta)X_{13}+\alpha[X,X_{3}]+\beta[X,X_{13}].
\]
Or using (\ref{XX}), (\ref{XX_13}), (\ref{XX_313}), we have
\[
12u_1X_1-12v_1X_2=X(\alpha)X_3+X(\beta)X_{13}+\alpha(-v_1X_1-12uu_1X_2)+\beta(-12u_1X_2).
\]
Equating the coefficients of the independent operators $X_1$ and $X_3,$ we obtain the system of equations
\begin{gather*}
X(\alpha)=0, \ 12u_1=-v_1\alpha.
\end{gather*}
The first equality implies that $\alpha=const.$ This contradicts the second equation.  Consequently, the operators $X_3, \ X_{13}$ and $X_{313}$ are linearly independent and  $dimL_3=1.$

The linear area $L_4$ is  generated by the operators $X_{1313},\ X_{2313}$ and $X_{3313}.$ Thus, it is easy to see that $X_{1313}=X_{2313}=0,$ and the commutator $X_{3313}$ has the form
\[
X_{3313}=-24v_1\frac{\partial}{\partial u_1}+288uu_1\frac{\partial}{\partial v_1}+\ldots.
\]
Using formulas (\ref{XX}), (\ref{XX_313}), we find
\begin{gather*}
[X,X_{3313}]=-[X_{313},[X,X_3]]+[X_3,[X,X_{313}]]=\\
=-[X_{313},-v_1X_1-12uu_1X_2]+[X_3,12u_1X_1-12v_1X_2],
\end{gather*}
hence
\begin{equation}
[X,X_{3313}]=24v_1X_1-288uu_1X_2-12u_1X_{13}.\label{XX_3313}
\end{equation}
Let show that the operators $X_3,\ X_{13}, \ X_{313}$ and $X_{3313}$ are linearly independent. Suppose to the contrary, that  $X_{3313}=\alpha X_{3}+\beta X_{13}+\delta X_{313},$ where $\alpha, \ \beta,\ \delta$  is a function of the variables  $u,v,u_1,v_1,u_2,v_2,\ldots.$ According to Lemma \ref{lem2} and formulas (\ref{XX}) - (\ref{XX_3313}), we have
\begin{gather*}
24v_1X_1-288uu_1X_2-12u_1X_{13}=X(\alpha)X_{3}+X(\beta)X_{13}+X(\delta)X_{313}+\\
+\alpha(-v_1X_1-12uu_1X_2)+\beta(-12u_1X_2)+\delta(12u_1X_1-12v_1X_2).
\end{gather*}
Equating coefficients of the operators $X_3, \ X_{313}, \ X_1$ and $X_{2}, \ X_{13}$ in the resulting correlation,  we have
\begin{gather*}
X(\alpha)=0, \ X(\delta)=0,\ \ 24v_1=-v_1\alpha+12u_1\delta,\\
-288uu_1=-12uu_1\alpha-12u_1\beta-12v_1\delta, \ -12u_1=X(\beta).
\end{gather*}
From the first three equations follows that  $\alpha=-24, \ \delta=0.$  Substituting the values $\alpha$ and $\delta$ in the fourth equation we find that $\beta=48u.$ This contradicts the condition $ -12u_1=X(\beta)$. Hence $dimL_4=1.$

The area $L_5$ is the linear span of commutators $X_{13313},\ X_{23313}, \ X_{33313},$ $[X_{13},X_{313}].$ Using the Jacobi identity and the formulas (\ref{XX_13}) - (\ref{XX_3313}), we can show that 
\[
[X,X_{13313}]=-288u_1X_2, \ [X,X_{23313}]=0, \ [X,[X_{13},X_{313}]]=-288u_1X_2.
\]
Then, according to Lemma \ref{lem2} and equality (\ref{XX_13}), we obtain
\[
X_{13313}=24X_{13}, \ X_{23313}=0, \ [X_{13},X_{313}]=24X_{13}.
\] 
For operator $X_{33313}$ we have
\begin{gather*}
[X,X_{33313}]=-[X_{3313},[X,X_3]]+[X_3,[X,X_{3313}]]=\\
=-[X_{3313},-v_1X_1-12uu_1X_2]+[X_3,24v_1X_1-288uu_1X_2-12u_1X_{13}].
\end{gather*}
Rearranging this equation, we obtain the equality
\[
[X,X_{33313}]=576uu_1X_1-576uv_1X_2-60v_1X_{13}-12u_1X_{313}.
\]
Using this equality we can obtain, that the operators $X_3,\ X_{13}, \ X_{313}, \ X_{3313}$ and $X_{33313}$  are linearly independent, and hence $dimL_5=1.$

Finally, we consider the area $L_6=L\left\langle X_{133313},\ X_{233313}, \ X_{333313}, \ [X_{13},X_{3313}]\right\rangle.$ Easy to check the validity of the following formulas
\begin{gather*}
X_{233313}=0,\ X_{133313}=48X_{313},\ [X_{13},X_{3313}]=24X_{313},\\
[X,X_{333313}]=1152uv_1X_1+6912u^2u_1X_2-1296uu_1X_{13}-120v_1X_{313}-12u_1X_{3313}.
\end{gather*}
As above, we can prove that the operators $X_3,\ X_{13}, \ X_{313},\ X_{3313}$ and $X_{333313}$ are linearly independent and $dimL_6=1.$

Thus, we have shown that 
\[
dimL_k=1, \ k=2,3,4,5,6; \ \ dim\sum^n_{i=1}L_i=8, \ n=6.
\]
Apparently, these formulas are valid for any $k$ and $n,$ ie
\[
dimL_k=1, \ k\geq2; \ \ dim\sum^n_{i=1}L_i=n+2,
\]
and $x-$ring of the system of equations \eqref{Ps012} is a ring of slow growth.

\section{Symmetries of Painlev\'e I equation}

In this section we calculate the higher symmetry of the system (\ref{Ps1}).

Consider $x-$symmetries of the form
\begin{gather*}
f=f(x,y,p,q,p_1,q_1,\ldots,p_n,q_n),\ \ g=g(x,y,p,q,p_1,q_1,\ldots,p_n,q_n),\\
(p_\tau=f,\ q_\tau=g).
\end{gather*}
We move from variables $x,y,p,q,p_1,q_1,\ldots,p_n,q_n,\ldots$ to  $x,$ $y,$ $p,$ $q,$ $\omega,$ $w,$ $\omega_1,$ $w_1,\ldots,$ $\omega_{n-1},$ $w_{n-1},\ldots$. Then the symmetries of $f$ and $g$ can be written as
\[
f=f(x,y,p,q,\omega,w,\omega_1,w_1,\ldots,\omega_{n-1},w_{n-1}),\ \ g=g(x,y,p,q,\omega,w,\omega_1,w_1,\ldots,\omega_{n-1},w_{n-1}),
\]
where $\omega$ and $w$ are $x-$integrals of the first order of equations (\ref{Ps1}) and $\omega_k=D^k\omega, \ w_k=D^kw, \ k=1,2,\ldots.$

The defining system of equations has the following view
\begin{equation}
D\bar{D}f=Dg,\ \ D\bar{D}g=12p_1Df.\label{opr_system}
\end{equation}

Further we introduce the notation
\begin{equation}\label{Df}
\begin{aligned}
Df=F(x,y,p,q,\omega,w,\omega_1,w_1,\ldots,\omega_{n},w_{n}), \\ 
Dg=G(x,y,p,q,\omega,w,\omega_1,w_1,\ldots,\omega_{n},w_{n}).
\end{aligned}
\end{equation}
Then equations \eqref{opr_system} become
\begin{equation}
\bar{D}F=G,\ \ \bar{D}G=12p_1F.\label{opr_system1}
\end{equation}

From \eqref{Df}, \eqref{opr_system1} we get that $F_p=F_q=G_p=G_q=0$ and, hence equations \eqref{opr_system1} are equivalent to the following
\begin{equation}
F_y=G,\ \ G_{y}=12p_1F.\label{opr_system2}
\end{equation}
Since equations \eqref{Ps1} have $x-$integrals of the first order, then there exists a function $h=h(y,\omega,w)$ such that $p_1=h.$ Thus from \eqref{opr_system2} we get
\begin{equation}
F_{yy}=12h(y,\omega,w)F.\label{h}
\end{equation}
Consider the symmetry of the first order 
\[
f=p_1, \ g=q_1.
\]
We have
\[
F=Df=Dh=h_{\omega}\omega_1+h_{w}w_1,
\]
and equation \eqref{h} becomes
\[
h_{\omega yy}\omega_1+h_{wyy}w_1=12h(h_{\omega}\omega_1+h_{w}w_1).
\]
Which implies that
\[
h_{\omega yy}=12hh_{\omega},\ \ h_{wyy}=12hh_{w}.
\]
Hence the functions $h_{\omega}$ and $h_{w}$ are particular solutions of \eqref{h}. We show that the solutions $h_{\omega}$ and $h_{w}$ are linearly independent. Indeed, if $h_{\omega}=c(\omega,w)h_w,$ then $h=h(y,\alpha(\omega,w))=h(y,W).$ And, therefore, system \eqref{Ps1} has an integral $W=W(y,p_1),$ which is impossible because of equality \eqref{w}.

Thus, the general solution of \eqref{h} has the form
\begin{equation}\label{F}
F=h_{\omega}A(x,\omega,w,\omega_1,w_1,\ldots,\omega_{n},w_{n})+h_{w}B(x,\omega,w,\omega_1,w_1,\ldots,\omega_{n},w_{n}).
\end{equation}
Finally from equations \eqref{Df}, \eqref{opr_system2},  \eqref{F}  it follows that the higher local $x-$Lie-B\"ucklund symmetries of \eqref{Ps1}  are given by
\begin{gather*}
f=D^{-1}(h_{\omega}A+h_{w}B), \ \ g=D^{-1}(h_{\omega y}A+h_{wy}B), 
\end{gather*}
where functions $A=A(x,\omega,w,\omega_1,w_1,\ldots,\omega_{n},w_{n})$ and $B=B(x,\omega,w,\omega_1,w_1,\ldots,\omega_{n},w_{n})$ satisfy
\[
\frac{\delta}{\delta \omega}(h_{\omega}A+h_{w}B)=\frac{\delta}{\delta w}(h_{\omega}A+h_{w}B)=0.
\]

Further we construct the $y-$symmetries 
\begin{gather*}
\varphi=\varphi(x,y,p,q,\bar{p}_1,\bar{q}_1,\ldots,\bar{p}_n,\bar{q}_n),\ \ \psi=\psi(x,y,p,q,\bar{p}_1,\bar{q}_1,\ldots,\bar{p}_m,\bar{q}_m),\\
(p_\tau=\varphi,\ q_\tau=\psi)
\end{gather*}
for system (\ref{Ps1}). 

Let the order of variables $p,q,\bar{p}_1,\bar{q}_1,\bar{p}_2,\bar{q}_2,\ldots$ the functions $D\varphi$ and $D\psi$ be equal to $n$ and $m$ correspondingly. Then, from the defining equations 
\begin{equation}\label{opr_system_y}
D\bar{D}\varphi=D\psi,\ \ D\bar{D}\psi=12p_1D\varphi
\end{equation}
it follows that $n+1=m$ and $m+1=n$  and hence
\[
D\varphi=F(x,y,p_1,q_1), \ \ D\psi=G(x,y,p_1,q_1).
\]
On the other hand
\begin{gather*}
D\varphi(x,y,p,q,\bar{p}_1,\bar{q}_1,\ldots,\bar{p}_n,\bar{q}_n)=\frac{\partial \varphi}{\partial x}+p_1\frac{\partial \varphi}{\partial p}+q_1\left(\frac{\partial }{\partial \bar{p}_1}+\frac{\partial}{\partial q}\right)\varphi+\\
+(6p^2_1+y)\left(\frac{\partial }{\partial \bar{p}_2}+\frac{\partial}{\partial \bar{q}_1}\right)\varphi+(12p_1q_1+1)\left(\frac{\partial }{\partial \bar{p}_3}+\frac{\partial}{\partial \bar{q}_2}\right)\varphi+\ldots.
\end{gather*}
Hence the function $\varphi$ satisfies the equations
\begin{gather*}
\frac{\partial \varphi}{\partial x}=\alpha(x,y), \ \frac{\partial \varphi}{\partial p}=\alpha _0(x,y), \ \left(\frac{\partial }{\partial \bar{p}_1}+\frac{\partial}{\partial q}\right)\varphi=\alpha _1(x,y),\\
\left(\frac{\partial }{\partial \bar{p}_2}+\frac{\partial}{\partial \bar{q}_1}\right)\varphi=\alpha _2(x,y),\ldots, \ \left(\frac{\partial }{\partial \bar{p}_{n}}+\frac{\partial}{\partial \bar{q}_{n-1}}\right)\varphi=\alpha _n(x,y).
\end{gather*}
It is easy to obtain from it, that
\begin{equation}\label{phi}
\varphi=\beta(x,y)+\beta _0(y)p+\beta _1(y)\bar{p}_1+\ldots+\beta _{n}(y)\bar{p}_n+h(y,\bar{\omega},\bar{\omega}_1,\ldots,\bar{\omega}_{n-1}),
\end{equation}
where $\bar{\omega}=\bar{p}_1-q$ is the $y-$integral of system (\ref{Ps1}).\\
An analogous formula holds true for the function $\psi:$
\begin{equation}\label{psi}
\psi=\gamma(x,y)+\gamma _0(y)p+\gamma _1(y)\bar{p}_1+\ldots+\gamma _{m}(y)\bar{p}_m+H(y,\bar{\omega},\bar{\omega}_1,\ldots,\bar{\omega}_{m-1}).
\end{equation}

Since the functions $\varphi=h(y,\bar{\omega},\bar{\omega}_1,\ldots,\bar{\omega}_{n-1})$ and $\psi=H(y,\bar{\omega},\bar{\omega}_1,\ldots,\bar{\omega}_{m-1})$ are define symmetries of the system (\ref{Ps1}) for any $h$ and $H,$ then we obtain from (\ref{phi}) and (\ref{psi}), that
\begin{equation}\label{psi_phi}
\varphi=\beta(x,y)+\sum_{k=0}^{n}\beta _k(y)\bar{p}_k,\ \ \psi=\gamma(x,y)+\sum_{k=0}^{m}\gamma _k(y)\bar{p}_k
\end{equation}
are also symmetries.
Substituting (\ref{psi_phi}) into defining system \eqref{opr_system_y}, we find, that
\[
\varphi=\varphi(y),\ \ \psi=\psi(y).
\]
Thus $y-$symmetries of the equations  (\ref{Ps1}) are calculated by formulae
\[
\varphi=h(y,\bar{\omega},\bar{\omega}_1,\ldots,\bar{\omega}_{n-1}), \ \ \psi=H(y,\bar{\omega},\bar{\omega}_1,\ldots,\bar{\omega}_{m-1}).
\]

\section{Symmetries of Painlev\'e III equation}

System of equations \eqref{main_eq3} corresponding to Painlev\'e III equation have the form
\begin{equation} \label{Ps3}
p_{xy}=q_x, \ \ yp_xq_{xy}=yq_{x}^2-p_xq_x+\delta y+\beta p_x+\alpha p_x^3 +\gamma yp_x^4.
\end{equation}
In this section higher symmetries for the system  (\ref{Ps3}) are built.

The $x-$characteristic Lie ring of system (\ref{Ps3}) is generated by vector fields 
\begin{gather*}
X_1=\frac{\partial}{\partial p}, \ \ X_2=\frac{\partial}{\partial q}, \\
X_3=\frac{\partial}{\partial y}+q_1\frac{\partial}{\partial p_1}+\left(\frac{q^2_1}{p_1}-\frac{q_1}{y}+\delta\frac{1}{p_1}+\beta\frac{1}{y}+\alpha\frac{p^2_1}{y}+\gamma p^3_1\right)\frac{\partial}{\partial q_1}+\ldots.
\end{gather*} 
Since the coefficients of the vector field $X_3$ is independent of $p$ and $q,$ the dimension of $x-$ring is equal to three and $x-$integrals $\omega=\omega(y,p_1,q_1) \ \mbox{and}\ w=w(y,p_1,q_1)$ are defined by the partial differential equation
\[
\left(\frac{\partial}{\partial y}+q_1\frac{\partial}{\partial p_1}+\left(\frac{q^2_1}{p_1}-\frac{q_1}{y}+\delta\frac{1}{p_1}+\beta\frac{1}{y}+\alpha\frac{p^2_1}{y}+\gamma p^3_1\right)\frac{\partial}{\partial q_1}\right)\Phi=0.
\]
Note, that $\omega =const$ and $w=const$ are define the integrals of the initial equation (\ref{penleve3}).  

Consider higher  $x-$ symmetries for equations (\ref{Ps3})
\begin{gather*}
f=f(x,y,p,q,p_1,q_1,\ldots,p_n,q_n),\ \ g=g(x,y,p,q,p_1,q_1,\ldots,p_n,q_n),\\
(p_\tau=f,\ q_\tau=g).
\end{gather*}
We move from variables $x,y,p,q,p_1,q_1,\ldots,p_n,q_n,\ldots$ to $x,$ $y,$ $p,$ $q,$ $\omega,$ $w,$ $\omega_1,$ $w_1,\ldots,$ $\omega_{n-1},$ $w_{n-1},\ldots$. Then the symmetry of $f$ and $g$ can be written as
\[
f=f(x,y,p,q,\omega,w,\omega_1,w_1,\ldots,\omega_{n-1},w_{n-1}),\ \ g=g(x,y,p,q,\omega,w,\omega_1,w_1,\ldots,\omega_{n-1},w_{n-1}).
\]

The defining system for system (\ref{Ps3}) has the form
\begin{equation}\label{opr_system3}
D\bar{D}f=Dg,\ \ D\bar{D}g=\left(2\frac{q_1}{p_1}-\frac{1}{y}\right)Dg+\left(-\frac{q^2_1}{p^2_1}-\delta\frac{1}{p^2_1}+2\alpha\frac{p_1}{y}+3\gamma p^2_1\right)Df,
\end{equation}
which can be written as
\begin{equation}\label{opr_systemP3}
\bar{D}F=G,\ \ \bar{D}G=\left(2\frac{q_1}{p_1}-\frac{1}{y}\right)G+\left(-\frac{q^2_1}{p^2_1}-\delta\frac{1}{p^2_1}+2\alpha\frac{p_1}{y}+3\gamma p^2_1\right)F,
\end{equation}
where 
\begin{equation}\label{Df3}
\begin{aligned}
Df=F(x,y,p,q,\omega,w,\omega_1,w_1,\ldots,\omega_{n},w_{n}), \\ 
Dg=G(x,y,p,q,\omega,w,\omega_1,w_1,\ldots,\omega_{n},w_{n}).
\end{aligned}
\end{equation}
Clearly that $F_p=F_q=G_p=G_q=0$ and the system \eqref{opr_systemP3} is equivalent to
\begin{equation}\label{opr_system33}
F_y=G,\ \ G_{y}=\left(2\frac{q_1}{p_1}-\frac{1}{y}\right)G+\left(-\frac{q^2_1}{p^2_1}-\delta\frac{1}{p^2_1}+2\alpha\frac{p_1}{y}+3\gamma p^2_1\right)F.
\end{equation}
Since the system \eqref{Ps3} has $x-$ integrals of the first order, then there are functions $h=h(y,\omega,w)$ and $H=H(y,\omega,w)$ such that $p_1=h, \ q_1=H.$ Then the second equation \eqref{opr_system33} can be written as
\begin{equation}\label{hH}
F_{yy}=\left(2\frac{H}{h}-\frac{1}{y}\right)F_y+\left(-\frac{H^2}{h^2}-\delta\frac{1}{h^2}+2\alpha\frac{h}{y}+3\gamma h^2\right)F.
\end{equation}
For the symmetry of the first order
\[
f=p_1, \ g=q_1
\]
equation \eqref{hH} is transformed as follows
\[
h_{\omega yy}\omega_1+h_{wyy}w_1=\left(2\frac{H}{h}-\frac{1}{y}\right)(h_{\omega y}\omega_1+h_{wy}w_1)+\left(-\frac{H^2}{h^2}-\delta\frac{1}{h^2}+2\alpha\frac{h}{y}+3\gamma h^2\right)(h_{\omega}\omega_1+h_{w}w_1),
\]
or
\begin{gather*}
h_{\omega yy}=\left(2\frac{H}{h}-\frac{1}{y}\right)h_{\omega y}+\left(-\frac{H^2}{h^2}-\delta\frac{1}{h^2}+2\alpha\frac{h}{y}+3\gamma h^2\right)h_{\omega},\\ 
h_{wyy}=\left(2\frac{H}{h}-\frac{1}{y}\right)h_{w y}+\left(-\frac{H^2}{h^2}-\delta\frac{1}{h^2}+2\alpha\frac{h}{y}+3\gamma h^2\right)h_{w}.
\end{gather*}
Hence the functions $h_{\omega}$ and $h_{w}$ are a particular solutions of \eqref{hH}.  It is easy to show that they are linearly independent, and so the general solution \eqref{hH} has the form
\begin{equation}\label{FF}
F=h_{\omega}A(x,\omega,w,\omega_1,w_1,\ldots,\omega_{n},w_{n})+h_{w}B(x,\omega,w,\omega_1,w_1,\ldots,\omega_{n},w_{n}).
\end{equation}
Finally, from equations \eqref{Df3}, \eqref{opr_system33},  \eqref{FF} it follows that the local higher $x-$symmetry Lie-B\"acklund of the system \eqref{Ps3} is defined from 
\begin{gather*}
f=D^{-1}(h_{\omega}A+h_{w}B), \ \ g=D^{-1}(h_{\omega y}A+h_{wy}B), 
\end{gather*}
where functions $A=A(x,\omega,w,\omega_1,w_1,\ldots,\omega_{n},w_{n})$ and $B=B(x,\omega,w,\omega_1,w_1,\ldots,\omega_{n},w_{n})$ satisfy
\[
\frac{\delta}{\delta \omega}(h_{\omega}A+h_{w}B)=\frac{\delta}{\delta w}(h_{\omega}A+h_{w}B)=0.
\]

The $y-$characteristic Lie ring of equations (\ref{Ps3}) is defined by the vector fields
\begin{gather*}
Y_1=\frac{\partial}{\partial p_1}, \ \ Y_2=\frac{\partial}{\partial q_1}, \\
Y_3=\frac{\partial}{\partial x}+p_1\frac{\partial}{\partial p}+q_1\frac{\partial}{\partial q}+q_1\frac{\partial}{\partial \bar{p}_1}+\left(\frac{q^2_1}{p_1}-\frac{q_1}{y}+\delta\frac{1}{p_1}+\beta\frac{1}{y}+\alpha\frac{p^2_1}{y}+\gamma p^3_1\right)\frac{\partial}{\partial \bar{q}_1}+\ldots.
\end{gather*}
It is easy to see that  $y-$ring is infinite. However, the system of equations (\ref{Ps3}) has $y-$integral of the first order $\bar{\omega}=\bar{p}_1-q.$ Define higher $y-$symmetries
\begin{gather*}
\varphi=\varphi(x,y,p,q,\bar{p}_1,\bar{q}_1,\ldots,\bar{p}_n,\bar{q}_n),\ \ \psi=\psi(x,y,p,q,\bar{p}_1,\bar{q}_1,\ldots,\bar{p}_m,\bar{q}_m),\\
(p_\tau=\varphi,\ q_\tau=\psi)
\end{gather*}
for equations (\ref{Ps3}). 

Let the order of variables $p,q,\bar{p}_1,\bar{q}_1,\bar{p}_2,\bar{q}_2,\ldots$ the functions $D\varphi$ and $D\psi$ be equal to $n$ and $m$ correspondingly. Then, from the defining system of equations 
\begin{equation}\label{opr_system_y3}
D\bar{D}\varphi=D\psi,\ \ D\bar{D}\psi=\left(2\frac{q_1}{p_1}-\frac{1}{y}\right)D\psi+\left(-\frac{q^2_1}{p^2_1}-\delta\frac{1}{p^2_1}+2\alpha\frac{p_1}{y}+3\gamma p^2_1\right)D\varphi
\end{equation}
it follows that $n+1=m$ and  $m+1=m,$ hence
\[
D\varphi=F(x,y,p_1,q_1), \ \ D\psi=G(x,y,p_1,q_1).
\]
On the other hand
\begin{gather*}
D\varphi(x,y,p,q,\bar{p}_1,\bar{q}_1,\ldots,\bar{p}_n,\bar{q}_n)=\frac{\partial \varphi}{\partial x}+p_1\frac{\partial \varphi}{\partial p}+q_1\left(\frac{\partial }{\partial \bar{p}_1}+\frac{\partial}{\partial q}\right)\varphi+\\
+\left(\frac{q^2_1}{p_1}-\frac{q_1}{y}+\delta\frac{1}{p_1}+\beta\frac{1}{y}+\alpha\frac{p^2_1}{y}+\gamma p^3_1\right)\left(\frac{\partial }{\partial \bar{p}_2}+\frac{\partial}{\partial \bar{q}_1}\right)\varphi+\ldots.
\end{gather*}
Therefore, the function $\varphi$  satisfies the system
\begin{gather*}
\frac{\partial \varphi}{\partial x}=\alpha(x,y), \ \frac{\partial \varphi}{\partial p}=\alpha _0(x,y), \ \left(\frac{\partial }{\partial \bar{p}_1}+\frac{\partial}{\partial q}\right)\varphi=\alpha _1(x,y),\\
\left(\frac{\partial }{\partial \bar{p}_2}+\frac{\partial}{\partial \bar{q}_1}\right)\varphi=\alpha _2(x,y),\ldots, \ \left(\frac{\partial }{\partial \bar{p}_{n}}+\frac{\partial}{\partial \bar{q}_{n-1}}\right)\varphi=\alpha _n(x,y).
\end{gather*}
It is easy to obtain from it, that
\begin{equation}\label{phi3}
\varphi=\beta(x,y)+\beta _0(y)p+\beta _1(y)\bar{p}_1+\ldots+\beta _{n}(y)\bar{p}_n+h(y,\bar{\omega},\bar{\omega}_1,\ldots,\bar{\omega}_{n-1}).
\end{equation}
An analogous formula holds true for the function $\psi:$
\begin{equation}\label{psi3}
\psi=\gamma(x,y)+\gamma _0(y)p+\gamma _1(y)\bar{p}_1+\ldots+\gamma _{m}(y)\bar{p}_m+H(y,\bar{\omega},\bar{\omega}_1,\ldots,\bar{\omega}_{m-1}).
\end{equation}

Since the functions $\varphi=h(y,\bar{\omega},\bar{\omega}_1,\ldots,\bar{\omega}_{n-1})$ and $\psi=H(y,\bar{\omega},\bar{\omega}_1,\ldots,\bar{\omega}_{m-1})$  are symmetries of the system (\ref{Ps3}) for any  $h$ and $H,$ then we obtain from (\ref{phi3}) and (\ref{psi3}), that
\begin{equation}\label{psi_phi3}
\varphi=\beta(x,y)+\sum_{k=0}^{n}\beta _k(y)\bar{p}_k,\ \ \psi=\gamma(x,y)+\sum_{k=0}^{m}\gamma _k(y)\bar{p}_k
\end{equation}
are also symmetries.
Substituting (\ref{psi_phi3}) into defining system \eqref{opr_system_y3}, we find that
\[
\varphi=\varphi(y),\ \ \psi=\psi(y).
\]
Thus, $y-$symmetry of the system of equations (\ref{Ps3})  are calculate by formulae
\[
\varphi=h(y,\bar{\omega},\bar{\omega}_1,\ldots,\bar{\omega}_{n-1}), \ \ \psi=H(y,\bar{\omega},\bar{\omega}_1,\ldots,\bar{\omega}_{m-1}).
\]

\subsection*{Acknowledgements}

One of authors A.V.Zhiber gratefully acknowledge financial support from a Russian Science Foundation grant (project 15-11-20007).


\pdfbookmark[1]{References}{ref}
\LastPageEnding

\end{document}